\documentclass[reprint, amsmath, amssymb, aps, prd]{revtex4-2}

\usepackage{graphicx}
\usepackage{bm}
\usepackage{hyperref}
\usepackage{physics}

\newcommand{\pvec}[1]{\vec{#1}}

\newcommand{\bham}{Institute for Gravitational Wave Astronomy \& School of Physics and Astronomy, University of Birmingham, Edgbaston, Birmingham B15 2TT, UK}

\begin{document}

\title{
Frequency-Domain Analysis of Black-Hole Ringdowns
}

\author{Eliot Finch}
\email{efinch@star.sr.bham.ac.uk}
\affiliation{\bham}

\author{Christopher J. Moore}
\email{cmoore@star.sr.bham.ac.uk}
\affiliation{\bham}

\date{\today}

\begin{abstract}
    We propose a novel, frequency-domain approach to the analysis of the gravitational-wave ringdown signal of binary black holes and the identification of quasinormal mode frequencies of the remnant.
    Our approach avoids the issues of spectral leakage that would normally be expected (associated with the abrupt start of the ringdown) by modeling the inspiral and merger parts of the signal using a flexible sum of sine-Gaussian wavelets truncated at the onset of the ringdown.
    Performing the analysis in the frequency domain allows us to use standard (and by now well-established) Bayesian inference pipelines for gravitational-wave data as well as giving us the ability to readily search over the sky position and the ringdown start time, although we find that it is necessary to use an informative prior for the latter.
    We test our method by using it to analyze several simulated signals with varying signal-to-noise ratios injected into two- and three-detector networks.
    We find that our frequency-domain approach is generally able to place tighter constraints on the remnant black-hole mass and spin than a standard time-domain analysis.
\end{abstract}

\maketitle

\section{Introduction}\label{sec:introduction}

The LIGO \cite{LIGOScientific:2014pky} and Virgo \cite{VIRGO:2014yos} observatories now routinely observe gravitational-wave (GW) signals from the inspiral, merger and ringdown of compact binaries \cite{LIGOScientific:2018mvr, LIGOScientific:2020ibl}.
Most of these GW signals come from binary black holes (BHs), and those with the highest masses [say, with detector frame total masses in the range $\sim (50$ -- $500)\,M_\odot$] typically exhibit loud ringdown signals that are sometimes visible in the whitened strain data.

The ringdown is associated with the system settling down into its final, stationary state.
Within general relativity (GR), the remnant is generally assumed to be a Kerr BH which is fully described by only a mass and spin (i.e.\ the \emph{no-hair} theorem).
As the merger and ringdown proceeds, the GW amplitude decreases and the final stages of this process can be well-described as linear perturbations of a remnant Kerr BH.
Perturbation theory identifies a discrete spectrum of complex (i.e.\ damped) frequencies $\omega_{\ell m n}$ which are prominent in the ringdown \cite{Berti:2009kk}.
These oscillations, known as quasinormal modes (QNMs) occur in pairs (``regular'' and ``mirror'' modes) and are indexed by integers $\ell \geq 2$ and $-\ell \leq m \leq \ell$ (spherical harmonic indices) and $n \geq 0$ (overtone index).
Hereafter, we use the term ringdown to mean the part of the signal that can be described by a superposition of QNMs.

Recent theoretical studies using catalogs of numerical relativity binary BH simulations suggest the ringdown typically starts early in the merger process, i.e.\ at or even slightly before the time of peak strain amplitude. 
This is only possible if the ringdown modeling includes overtones ($n \geq 1$) \cite{Giesler:2019uxc, JimenezForteza:2020cve, Forteza:2021wfq}, and possibly a combination of mirror modes and/or higher harmonics ($\ell\geq 3$) \cite{Cook:2020otn, Dhani:2020nik, Finch:2021iip}.
This early start time is good for the prospects of observing QNMs because it means the signal amplitude is still large when the ringdown starts and consequently the signal-to-noise ratio (SNR) in the ringdown is large.
Ref.~\cite{LIGOScientific:2020tif} was able to identify a QNM in 17 of the binary BHs observed so far, and furthermore found evidence for an overtone in two cases (namely GW150914 \cite{LIGOScientific:2016aoc} and GW190521\_074359).
We note that Ref.~\cite{LIGOScientific:2020tif} found no strong evidence for harmonics or overtones in the heaviest source, GW190521 \cite{LIGOScientific:2020iuh}.

Once QNMs have been correctly identified in an observed signal, they provide an exciting opportunity for testing GR, the Kerr metric hypothesis, and the no-hair theorem. 
The idea of using QNM frequencies for such tests, sometimes referred to as \emph{BH spectroscopy}, predates the detection of GWs \cite{Dreyer:2003bv, Berti:2005ys, Berti:2007zu, Berti:2016lat}.
Therefore, experimental QNM tests of GR started immediately with the first GW observation; Ref.~\cite{LIGOScientific:2016lio} found that the data following the peak of GW150914 was consistent with the least-damped QNM of the expected remnant.
Subsequently, several groups reanalyzed the ringdown of GW150914 with the aim of identifying additional QNMs for use in spectroscopic tests \cite{Carullo:2019flw, Isi:2019aib, Brito:2018rfr}. 
With the second GW catalog, similar analyses are now routinely performed on all suitable events \cite{LIGOScientific:2020tif}.
Besides BH spectroscopy, other types of QNM test are possible; for example, Ref.~\cite{Isi:2020tac} used the QNM frequencies of GW150914 to measure the horizon area of the remnant and thereby test Hawking's area theorem \cite{Hawking:1971tu}.
The applications mentioned so far use only the QNM frequencies; however, the excitation amplitudes and phases of the QNMs also carry useful information about the progenitor binary (see, for example, Refs.~\cite{Hughes:2019zmt, Berti:2006wq}). 

Unfortunately, QNMs are difficult to work with both from a data analysis and a theoretical perspective. 
The start time is uncertain; even with clean, noise-free numerical simulations an unambiguous determination of the ringdown start time, $t_0$, is impossible (see, for example, Ref.~\cite{Thrane:2017lqn}). 
The start of the ringdown is also abrupt in the time domain, which makes it non-local in the frequency domain and is the source of spectral leakage problems. 
Theoretically, the QNM content is uncertain; \emph{a priori} it is not known which modes should be included in the analysis.
The mode excitations depend on the initial conditions of the system in a non-trivial way, and which QNMs are detectable is also a function of the chosen ringdown start time and the SNR.
Also, QNMs do not form a complete basis \cite{Berti:2009kk}; the late-time signal contains additional components that decay more slowly, known as \emph{tails}.
Finally, it is also known that very similar compact objects can nevertheless have completely different QNM spectra \cite{Nollert:1996rf}.

Despite the difficulties, detecting and characterizing QNMs is a key goal in GW astronomy. 
The most natural approach to deal with the abrupt ringdown start is to work in the time domain. 
This differs from other GW data analysis which is almost universally performed in the frequency domain. 
However, by working in the time domain, the data can be cut precisely at a chosen $t_0$ and an analysis performed on only the ringdown part of the signal, $t \geq t_0$, without any spectral leakage.
As the particular segment of data to be analyzed is chosen before the analysis begins, the ringdown start time (and consequently the source sky location) usually must be fixed in these analyses. 
Although, see Ref.~\cite{Carullo:2019flw} where a posterior on ringdown start time is obtained for GW150914 and Ref.~\cite{Isi:2021iql} where there is a discussion about how it would be possible, in principle, to vary the sky location.
Another drawback of working in the time domain is that the noise covariance matrix is no longer diagonal, increasing computational cost of the likelihood. 
The covariance matrix is constructed with the autocovariance function, which characterizes the noise in the time domain.
And, as discussed in Ref.~\cite{Isi:2021iql}, care must be taken when estimating the autocovariance to avoid corrupting the ringdown data. Therefore, there are additional subtleties in a time-domain analysis compared to a frequency-domain analysis.

Recently, in Ref.~\cite{Capano:2021etf}, an alternative approach to ringdown analysis was presented and applied to GW190521 where a higher harmonic was identified, apparently in contradiction with the results of Ref.~\cite{LIGOScientific:2020tif} (but there are important differences in the analyses). 
Although this alternative approach is expressed in the frequency domain, it uses a modified expression for the likelihood (involving \emph{in-painting} the data before the start of the ringdown in such a way as to remove the contribution to the likelihood) and it has been shown to be equivalent to the standard time-domain approach \cite{Isi:2021iql}.
That these two formally equivalent analyses \cite{LIGOScientific:2020tif, Capano:2021etf} can come to different conclusions regarding the QNM content of GW190521 highlights some of the difficulties that come with this type of analysis, where important choices (that can affect the result) for the ringdown start time have to be made and care must be taken with the noise estimation.

In this paper, we present a new approach to performing ringdown analyses in the frequency domain. 
We employ a flexible sum of truncated wavelets to model the inspiral-merger signal (inspired by \texttt{BayesWave} \cite{Cornish:2014kda, Cornish:2020dwh}) and QNMs to model the ringdown.
The general idea behind our approach is illustrated in Fig.~\ref{fig:demo}.
By working in the frequency domain, we can use the standard, and now very mature, GW data analysis pipelines (our analysis pipeline is built on the public \texttt{Bilby} package \cite{Ashton:2018jfp}).
Also, working in the frequency domain makes it trivial to search and marginalize over the sky position the ringdown start time, $t_0$. 
We hope this approach can complement existing time-domain analyses.

The details of our method are described in Sec.~\ref{sec:methods}, where we compare and contrast the time- and frequency-domain likelihood functions before introducing our frequency-domain approach.
In Sec.~\ref{sec:injection_study} we present the results of a series of analyses on simulated GW signals where we test the performance of our approach and compare it with time-domain methods.
We present our conclusions in Sec.~\ref{sec:discussion}.
A complete set of posterior samples from this work are made available at Ref.~\cite{finch_eliot_2021_5569759}.

\section{Methods}\label{sec:methods}

This section describes the details of the proposed frequency-domain approach to the analysis of BH ringdowns. 
Sec.~\ref{subsec:data_analysis} describes the GW likelihood function and its implementation in both the time and frequency domains.
Sec.~\ref{subsec:motivation} motivates our frequency-domain approach by describing an extreme limit in which it becomes equivalent to the standard, time-domain approach.
Finally, Sec.~\ref{subsec:model} describes the combination of truncated wavelets and QNMs that comprise our waveform model.

\subsection{Time- and Frequency-Domain Likelihoods}\label{subsec:data_analysis}

\begin{figure*}[t]
    \centering
    \includegraphics[width=2\columnwidth]{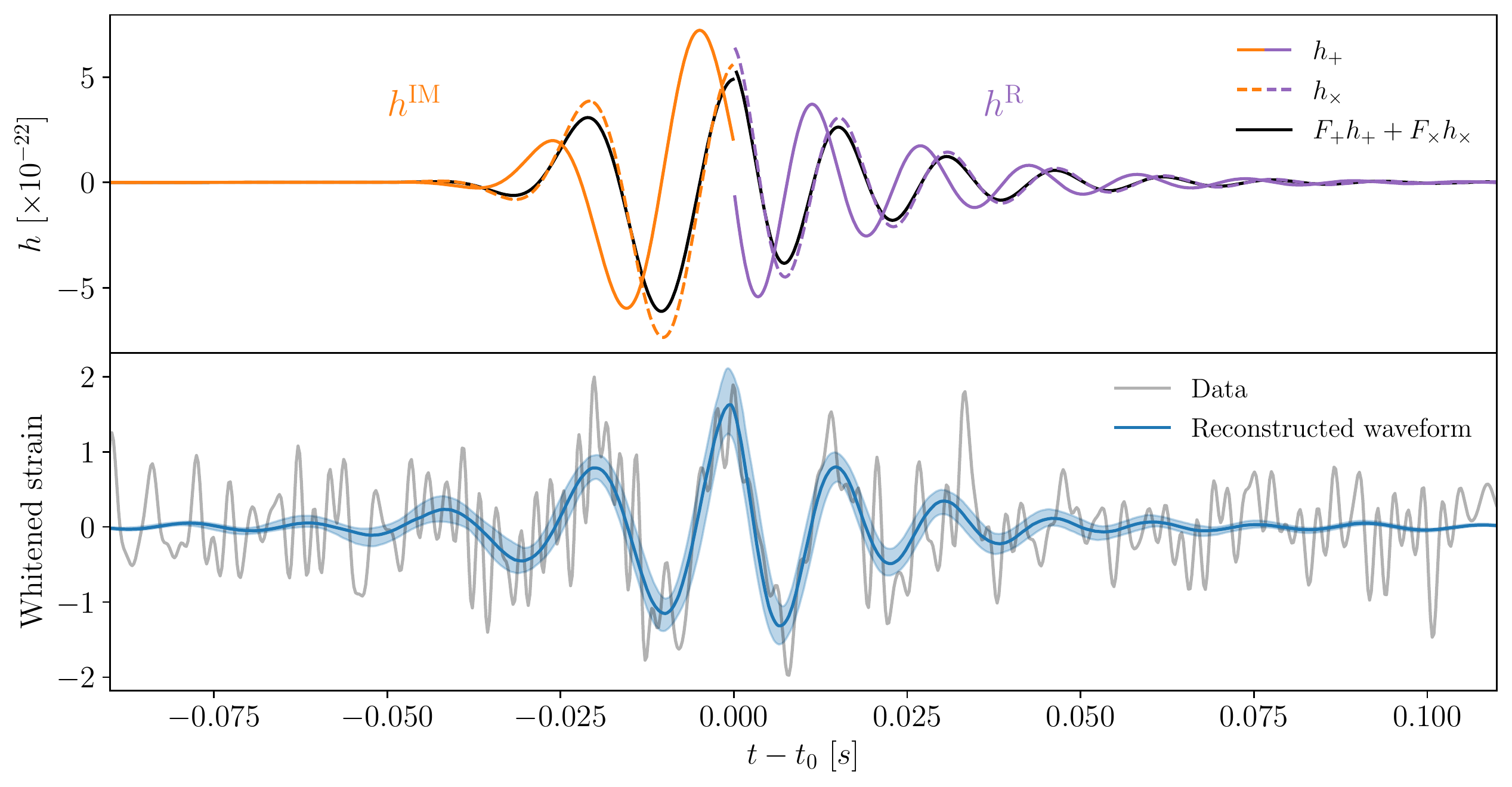}
    \caption{ \label{fig:demo}
        This figure is intended to illustrate the idea behind our approach for analyzing the ringdown in the frequency domain.
        The gray line in the bottom panel shows the whitened and band-passed Livingston time-series data around GW190521; this loud GW signal comes from a high-mass binary BH merger and exhibits a clear ringdown.
        For the purpose of illustration, we use the simplest version of our model where a single, truncated sine-Gaussian wavelet is used to model the inspiral-merger part of the signal and the fundamental $\ell=m=2,\ n=0$ QNM is used to model the ringdown.
        The ringdown start time, $t_0$, is allowed to vary as part of a Bayesian analysis which also searches over different values of wavelet parameters, the source sky position, the QNM amplitude and phase, and over the remnant BH mass and spin. Full details of this analysis will be presented elsewhere.
        The top panel shows the maximum likelihood waveform broken down into its wavelet ($h^{\rm IM}$, orange) and ringdown ($h^{\rm R}$, purple) parts as well as into plus (solid) and cross (dashed) polarizations. The discontinuity in our model can be clearly seen in the colored lines. However, when these polarizations are combined and projected [see Eq.~(\ref{eq:projection_antenna})] onto the interferometer (black line) the result is nearly continuous; we emphasize that this continuity has not been imposed by the model but is rather ``learnt'' from the data. When the projected maximum likelihood waveform is whitened according to the detector noise curve it becomes completely continuous, this is plotted as the central blue line in the bottom panel.
        In the bottom panel we also plot the uncertainty (90\% credible region; blue shaded band) on the recovered signal.
    }
\end{figure*}

Most GW data analysis is done in the frequency domain because, with the usual assumptions of stationary zero-mean Gaussian noise, the instrumental noise is fully described by the (one-sided) noise power spectral density (PSD), $S_n(f)$.
In the literature, the most commonly encountered expression for the log-likelihood in one interferometer is the integral
\begin{align} \label{eq:logL_FD_continuous}
    \log\mathcal{L}(d|\pvec{\theta}) = -2\int_{0}^{\infty}\!\mathrm{d}f\;\frac{|\tilde{d}(f)-\tilde{h}(f;\pvec{\theta})|^2}{S_n(f)} + \mathrm{norm},
\end{align}
where $d(t)$ is the observed data, and $h(t;\pvec{\theta})$ is the signal model (projected onto the interferometer) described by parameters $\pvec{\theta}$.
The normalization constant in the likelihood is unimportant for our purposes and will be dropped in all following equations.
A tilde denotes the Fourier transform of a time series.
Because the noise is uncorrelated between two well separated interferometers, the log-likelihood for a network is obtained by summing the independent contributions from each instrument.

In practice, it is necessary to work with discretely sampled time series; $d_j = d(t_j)$, where $t_j=j\delta t$ for $j=0,1,\ldots, J-1$, and where $1/\delta t$ is the sampling frequency.
The discrete Fourier transform $\tilde{d}_k = \tilde{d}(f_k)$ is sampled at (positive) frequencies $f_k=k/(J\delta t)$ for $k=0,1,\ldots,K-1$, where $K=\left \lfloor (J+2)/2\right \rfloor$.
The log-likelihood in terms of the discretely sampled frequency series is given by the following sum,
\begin{align} \label{eq:logL_FD_discrete}
    \log\mathcal{L}(d|\pvec{\theta}) = \frac{-2}{J\delta t}\sum_{k}\frac{|\tilde{d}_k-\tilde{h}_k(\pvec{\theta})|^2}{S_n(f_k)},
\end{align}
which can be compared to Eq.~(\ref{eq:logL_FD_continuous}).
The noise PSD is usually estimated from off-source data using a Welch periodogram \cite{1161901}.
The frequency-domain expression for the log-likelihood involves a single sum; the noise covariance matrix is diagonal in the frequency domain.
Although, for finite duration time series there can be small correlations between frequency bins \cite{Talbot:2021igi}.

The log-likelihood can also be expressed in the time domain via
\begin{align} \label{eq:logL_TD_discrete}
    \log\mathcal{L}(d|\pvec{\theta}) = -\frac{1}{2}\sum_{jj'}\qty[d_j-h_j(\pvec{\theta})] C^{-1}_{jj'} \qty[d_{j'}-h_{j'}(\pvec{\theta})],
\end{align}
where $C_{jj'}$ is the noise covariance matrix.
The time-domain expression for the log-likelihood involves a double sum over a dense covariance matrix which is computationally more costly to evaluate than the frequency-domain expression [$\mathcal{O}(J^2)$ as opposed to $\mathcal{O}(J)$].

Because the noise is assumed to be stationary, $C_{jj'}$ has the Toeplitz structure
\begin{align} \label{eq:Toeplitz}
    C_{jj'} = \rho_{|j-j'|},
\end{align} 
where $\rho_j$ is the noise autocovariance. 
This can also be estimated from off-source data using the following two-point expectation:
\begin{align} \label{eq:autocovariance}
    \rho_j = \frac{1}{\mathcal{J}}\sum_{j'=0}^{\mathcal{J}-1}n_{j'}n_{(j'+j)}.
\end{align}
Here, $\mathcal{J}$ is the length of some off-source data segment, which is usually chosen to be longer than the analysis data (i.e.\ $\mathcal{J} \gg J$) \cite{Isi:2021iql}.
It is also necessary to treat the ``edges'' of the data segment (i.e.\ where $j + j' > \mathcal{J}-1$) either by zero-padding or imposing periodicity: $\rho_j = \rho_{\mathcal{J}-j}$.
Although these different treatments result in an autocovariance that differs for large $j$, if $\mathcal{J}$ is sufficiently large then the autocovariance will be consistent for $j < J$ (which is what enters the calculation of the likelihood).

The two expressions for the log-likelihood are equivalent.
The noise autocovariance (which appears in the time-domain log-likelihood) is related to the PSD (in the frequency-domain log-likelihood) via a discrete Fourier transform (Wiener-Khinchin theorem), when imposing the circularity condition \cite{Isi:2021iql}:
\begin{align} \label{eq:WKtheorem}
    \frac{1}{2}S_n(f_k) = \delta t \sum_{j}\rho_j \exp\left(\frac{-2\pi ijk}{J}\right).
\end{align}
We use the inverse of Eq.~(\ref{eq:WKtheorem}) to estimate the autocovariance, which comes with the requirement that the off-source segment length used in the PSD estimate is much longer than the analysis length \cite{Isi:2021iql}.

The time-domain expression for the log-likelihood has hitherto been considered more suitable for ringdown analyses.
This is because in the time-domain expression no Fourier transform of the data or model is required, and so no periodicity has to be ensured.
The abrupt start of ringdown models means they do not satisfy this periodicity condition, which leads to spectral leakage upon Fourier transforming. 
When performing Fourier transforms of the GW data, periodicity is ensured by applying window functions to taper the data.
This makes it difficult to isolate the ringdown region of a GW signal in some data; a sharp cut at the ringdown start time would introduce a discontinuity, whereas a smooth window would either suppress the ringdown signal or risk contamination of the ringdown by including unwanted parts of the inspiral-merger (see Fig.~7 in \cite{Isi:2021iql} for an illustration of this). These problems are naturally avoided in the time domain. 
In the following sections we describe how these problems can also be overcome in a frequency-domain analysis.

\subsection{Marginalizing Over the Inspiral-Merger}\label{subsec:motivation}

We now motivate our approach to analyzing the ringdown by first discussing a special case in which it becomes formally equivalent to the standard time-domain approach. 

Consider first the case of a single interferometer.
The observed data is a discretely sampled time series:
\begin{align}
    \ldots,\ d_{-2},\ d_{-1},\ d_{0},\ d_{1},\ d_{2},\ \ldots
\end{align}
We assume that the ringdown has been identified as starting at the time of $d_0$.
The GW signal has a large amplitude around $d_0$, but decays to zero at early and late times.

The standard approach to analyzing the ringdown is to cut the data at the start time where the signal amplitude is large and take only the data after that time (i.e.\ $d_{0},\ d_{1},\ d_{2},\ \ldots$) and to model this using a superposition of QNMs described by parameters $\pvec{\theta}$ (e.g.\ the remnant mass, spin, amplitudes and phases for each QNM).
The likelihood is written as
\begin{align} \label{eq:motivation_A}
    \mathcal{L}(d_{0}, d_{1}, d_{2}, \ldots|\pvec{\theta}),
\end{align}
and, because we have cut the signal where the amplitude is large, this must be expressed in the time domain [Eq.~(\ref{eq:logL_TD_discrete})] to avoid problems of spectral leakage.

We could instead extend our analysis segment backwards by starting at $d_{-1}$, and then analyze this longer data stream with a new model that treats the value of the signal at $d_{-1}$ as being a completely free parameter. 
The model is otherwise unchanged at later times and is now described by parameters $(\hat{d}_{-1}, \pvec{\theta})$.
Note that the new model for $d_{-1}$ is entirely unphysical and is also discontinuous in the sense that there is no requirement that the model takes similar values at times $-1$ and 0.
The likelihood for this new model
\begin{align} \label{eq:motivation_B}
    \mathcal{L}(d_{-1}, d_{0}, d_{1}, d_{2}, \ldots|\hat{d}_{-1}, \pvec{\theta}),
\end{align} 
is also given by Eq.~(\ref{eq:logL_TD_discrete}), only with a slightly larger covariance matrix.
If we marginalize this with respect to the ``inspiral-merger parameter'' $\hat{d}_{-1}$ 
(adopting a flat, improper prior on $\hat{d}_{-1}$ ranging between $\pm\infty$)
then we recover the original likelihood in Eq.~(\ref{eq:motivation_A}), i.e.\
\begin{align} \label{eq:motivation_C}
    \mathcal{L}(d_{0}, d_{1}, d_{2}, &\ldots|\pvec{\theta}) = \\ &\int_{-\infty}^{\infty} \mathrm{d}\hat{d}_{-1}\; \mathcal{L}(d_{-1}, d_{0}, d_{1}, d_{2}, \ldots|\hat{d}_{-1}, \pvec{\theta}). \nonumber
\end{align}
This follows from well-known properties of the Gaussian distribution.

We can of course include more early-time data in a similar way. 
If we treat the value of the GW signal at each of the times $d_{-2},\ d_{-3},\ \ldots$ as free parameters and marginalize over all of them then we recover the original likelihood in Eq.~(\ref{eq:motivation_A}).
The point of doing this is that it allows us to start the analysis at early times when the amplitude of the GW signal in the data is small.
This means we can apply windowing, aka tapering, operations to the data without fear of suppressing the signal, and we can therefore transform the likelihood into the frequency domain without encountering spectral leakage problems.

The extension of this argument to multiple interferometers is straightforward. 
The values of the model at each early time in each interferometer must all be treated independently as free parameters and marginalized over in the manner of Eq.~(\ref{eq:motivation_C}).

The point we wish to emphasize is that an analysis of only the ringdown data (usually done in the time domain to avoid problems of spectral leakage) is equivalent to an analysis of all the data (which can be done in the frequency domain using standard GW data analysis techniques) if the inspiral-merger part of the signal is suitably marginalized out. 
This requires the use of an unphysical and discontinuous model for the inspiral-merger signal.
For the equivalence to be exact, the inspiral-merger model should include an extremely large number of free parameters (one for each time stamp in each interferometer), but we will argue below that in practice it is sufficient to use a smaller number of parameters provided a sufficiently flexible model is used.

This discussion motivates the model we describe in Sec.~\ref{subsec:model}. Once the likelihood is expressed in the frequency domain, several extensions of the analysis (such as treating the ringdown start time as a parameter of the model; see Sec.~\ref{subsec:t0}) become natural.

\subsection{Model}\label{subsec:model}

As it is clearly impractical to model the data at each time stamp as a free parameter, we propose to use a continuous (but very flexible) inspiral-merger model instead. 
We choose a sum of sine-Gaussian wavelets, which are then truncated at the ringdown start time and attached to a ringdown QNM model. 
With this method, we aim to model the full inspiral-merger-ringdown signal.

The ringdown model is zero for early times, and after a start time $t_0$ takes the form
\begin{align}
    h^\mathrm{R}(t) &= h_+^\mathrm{R}(t) - ih_\times^\mathrm{R}(t) \nonumber \\
    &= \sum_{\ell m n} A_{\ell m n} e^{-i[\omega_{\ell m n}(t-t_0) - \phi_{\ell m n}]}, \quad t \geq t_0,
\end{align}
where the complex QNM frequencies $\omega_{\ell m n} = 2\pi f_{\ell m n} - i/\tau_{\ell m n}$ are functions of the remnant BH mass $M_f$ and dimensionless spin magnitude $\chi_f$. Here, $f_{\ell m n}$ is the oscillation frequency, and $\tau_{\ell m n}$ is the damping time.
Each QNM is further described by an amplitude, $A_{\ell m n}$, and phase parameter, $\phi_{\ell m n}$. 
It is possible to analytically take the Fourier transform of this expression and thereby write the ringdown model in the frequency domain as
\begin{align}
    \Tilde{h}^{\mathrm{R}}(f) &= \int_{-\infty}^\infty \dd{t} \qty[h_+^\mathrm{R}(t) - ih_\times^\mathrm{R}(t)] e^{-2\pi i f t} \nonumber \\
    &= \sum_{\ell m n} \frac{A_{\ell m n} e^{-i[2\pi ft_0 - \phi_{\ell m n}]}}{i(\omega_{\ell m n} + 2\pi f)}.
\end{align}

We model the inspiral-merger part of the signal as a truncated sum of $W$ wavelets.
The inspiral-merger model is zero for late times, but before the start time $t_0$ takes the form
\begin{align}
    h^\mathrm{IM}(t) &=  h_+^\mathrm{IM}(t) - ih_\times^\mathrm{IM}(t) \nonumber \\
    &= \sum_{w=1}^{W} \mathcal{A}_w \exp \Bigg[-2\pi i \nu_w(t-\eta_w) \\
    &\hspace{2.46cm} - \qty(\frac{t-\eta_w}{\tau_w})^2 + i\varphi_w \Bigg], \quad t < t_0. \nonumber
\end{align}
The limit on time is equivalent to a model which is multiplied by a Heaviside step function, $H(t_0 - t)$. Each wavelet is described by five parameters: $\mathcal{A}_w$ and $\varphi_w$ are the wavelet amplitudes and phases, $\tau_w$ are the wavelet widths, $\nu_w$ are the wavelet frequencies, and $\eta_w$ are the wavelet central times. Again, it is possible to analytically take the Fourier transform of this expression and thereby write the inspiral-merger model in the frequency domain as
\begin{align}
    &\Tilde{h}^\mathrm{IM}(f) = \int_{-\infty}^\infty \dd{t} \qty[h_+^\mathrm{IM}(t) - ih_\times^\mathrm{IM}(t)] e^{-2\pi i f t} \\
    &\quad= \sum_{w=1}^{W} \mathcal{A}_w \exp[-2\pi i\nu_w \eta_w -\pi^2(f+\nu_w)^2\tau_w^2 +i\varphi] \nonumber\\
    &\quad\quad\times \frac{\sqrt{\pi}}{2}\tau_w \left(1 + \mathrm{erf} \left[ \frac{t_0-\eta_w}{\tau_w} + \pi i(f+\eta_w)\tau_w \right] \right).\nonumber
\end{align}

The full IMR model is simply given by
\begin{equation}
    h(t) = h^\mathrm{IM}(t) + h^\mathrm{R}(t).
\end{equation}
We refer the reader to Fig.~\ref{fig:demo}, which provides an illustration of this model with a single QNM and a single wavelet ($W=1$).

If $N$ QNMs are used, then the ringdown part of the model is described by $2N+2$ parameters.
The inspiral-merger part of the model is described by $5W$ parameters (these numbers do not include $t_0$).
Additionally, there are two sky position angles ($\alpha$ and $\delta$) and a polarization angle ($\psi$) which enter the detector response described below.
Finally, the ringdown start time ($t_0$) can also be treated as a model parameter in our frequency-domain approach.
For a typical choice of these parameters, our IMR model is discontinuous at $t_0$.
The inspiral-merger part of the model contains no information of the physics of the source and no attempt is made to enforce continuity between the two parts; this helps to decouple the ringdown inference from the inspiral-merger parts of the data and thereby ensure that we are really performing a ringdown analysis. 

To complete the description of our model, the detector response is given by projecting the waveform polarizations onto each interferometer (IFO) with the appropriate antenna patterns, $F^\mathrm{IFO}_{+,\times}$.
For a given sky location and GW polarization angle the detector response for each ${\mathrm{IFO}\in \{\mathrm{H}, \mathrm{L}, \mathrm{V} \}}$ is given by
\begin{align} \label{eq:projection_antenna}
    h^\mathrm{IFO}(t) = F^\mathrm{IFO}_+(\alpha, \delta, \psi) ~ &h_+(t + \Delta t_\mathrm{IFO}) \nonumber \\
    + F^\mathrm{IFO}_\times(\alpha, \delta, \psi) ~ &h_\times(t + \Delta t_\mathrm{IFO}).
\end{align}
Here, $\Delta t_\mathrm{IFO}(\alpha, \delta)$ accounts for the different signal arrival times at the detectors and is also a function of sky location.
By definition, $h_+(t) = \Re\{ h(t) \}$, and $h_\times(t) = -\Im \{ h(t) \}$.
Note, the frequency-domain waveforms presented here are Fourier transforms of the complex polarization sum $h_+(t) - ih_\times (t)$. This means the separation into the plus and cross polarizations is not as simple as for the time-domain waveforms. Instead, the property that $\Tilde{h}^*_{+,\times}(-f) = \Tilde{h}_{+,\times}(f)$ for a real time-series implies that
\begin{gather}
    \Tilde{h}_+(f) = \frac{\Tilde{h}(f) + \Tilde{h}^*(-f)}{2}, \\ 
    \Tilde{h}_\times(f) = -\frac{\Tilde{h}(f) - \Tilde{h}^*(-f)}{2i}.
\end{gather}

\section{Injection study}\label{sec:injection_study}

We use the numerical relativity surrogate NRHybSur3dq8 \cite{Varma:2018mmi} to simulate the full inspiral, merger and ringdown signal from GW190521-like and GW150914-like sources. 
We use these two sources to test our frequency-domain approach on the analysis of the ringdown and compare the results with a standard, time-domain analysis.
Results from the GW190521-like analyses are shown here while the results from the GW150914-like analyses are shown in Appendix~\ref{app:GW150914}.

For all the GW190521-like injections, the surrogate was initialized with a total mass of $271\,M_\odot$ (all masses are given in the detector frame) and a mass ratio of $1.27$. 
For simplicity, all of the component spins were set to zero and the inclination angle was also chosen to be zero (i.e.\ the source was injected ``face-on'').
The simulated sky location and GW polarization angle were taken to be the maximum likelihood values from the NRSur7dq4 analysis in Refs.~\cite{LIGOScientific:2020ibl, gwtc2datarelease} ($\alpha = 0.164$, $\delta = -1.14$, $\psi = 2.38$).
The distance to the binary was chosen so that it gives a particular value of the optimal SNR in Livingston; this was usually chosen to be 15 (corresponding to a distance of $4016\,\mathrm{Mpc}$) so that it would likely be possible to detect multiple QNMs (in particular overtones), but several smaller values are also considered in Sec.~\ref{subsec:overtones}.
We perform zero-noise injections (i.e. analyzing simulated data with a noise realization of zero) into a three-interferometer H-L-V LIGO-Virgo network (except in Sec.~\ref{subsec:t0} where a two-interferometer H-L injection is performed for comparison) and use the average PSDs from the first three months of O3 (available at Ref.~\cite{o3psd}).

\begin{figure}[t]
    \centering
    \includegraphics[width=\columnwidth]{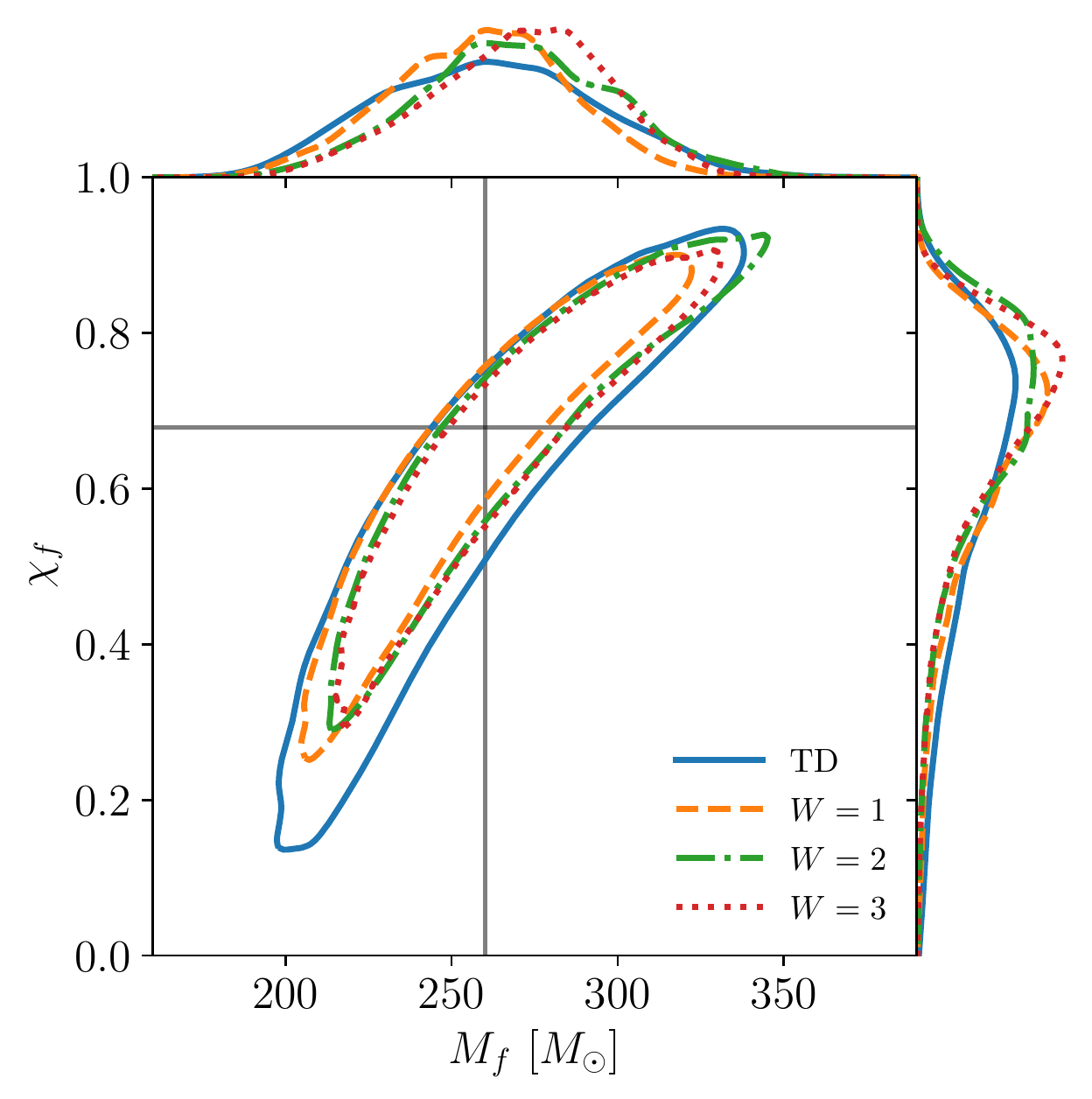}
    \caption{ \label{fig:mass_spin_corner_fixed_sky}
    Posteriors on the (detector frame) remnant mass and dimensionless spin for the GW190521-like injection using a single QNM and analyzed with a fixed sky position, GW polarization angle and ringdown start time.
    The main panel shows the $90\%$ confidence contour while the side plots show the one-dimensional marginalized posteriors.
    The solid blue line shows the results of a time-domain (TD) analysis.
    The other dashed and dotted lines show the results of frequency-domain analyses using different numbers of wavelets, $W$.
    The vertical and horizontal lines indicate the true values.
    }
\end{figure}

Although an important advantage of our approach is that it allows for easy marginalization over the source sky position and ringdown start time, we first apply it to the case where these are fixed. 
This allows us to compare our results more directly to those from a time-domain analysis.
The sky position and polarization angles $\alpha$, $\delta$ and $\psi$ are fixed to their injected values and the ringdown start time $t_0$ is fixed to be $12.7\,\mathrm{ms}$ ($\sim 10\,M_f$ in geometric units) after the peak of the strain (this choice follows the analysis of the real GW190521 signal in Ref.~\cite{LIGOScientific:2020iuh}).

\begin{figure*}
    \centering
    \begin{minipage}{0.49\linewidth}
    \includegraphics[width=0.9\textwidth]{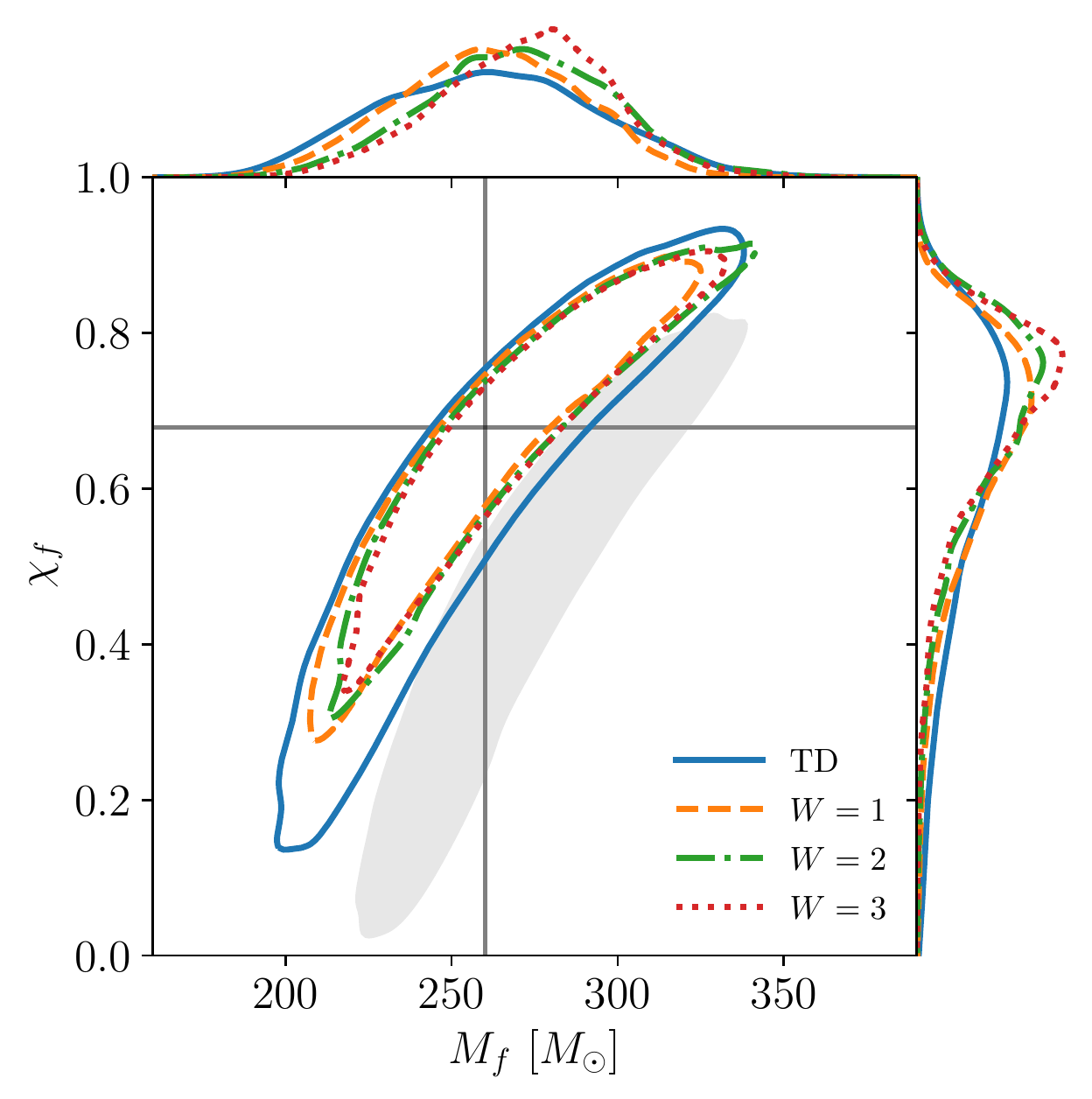} 
    \end{minipage}
    \hfill
    \begin{minipage}{0.49\linewidth}
    \vspace{1.2cm}
    \includegraphics[width=0.9\textwidth]{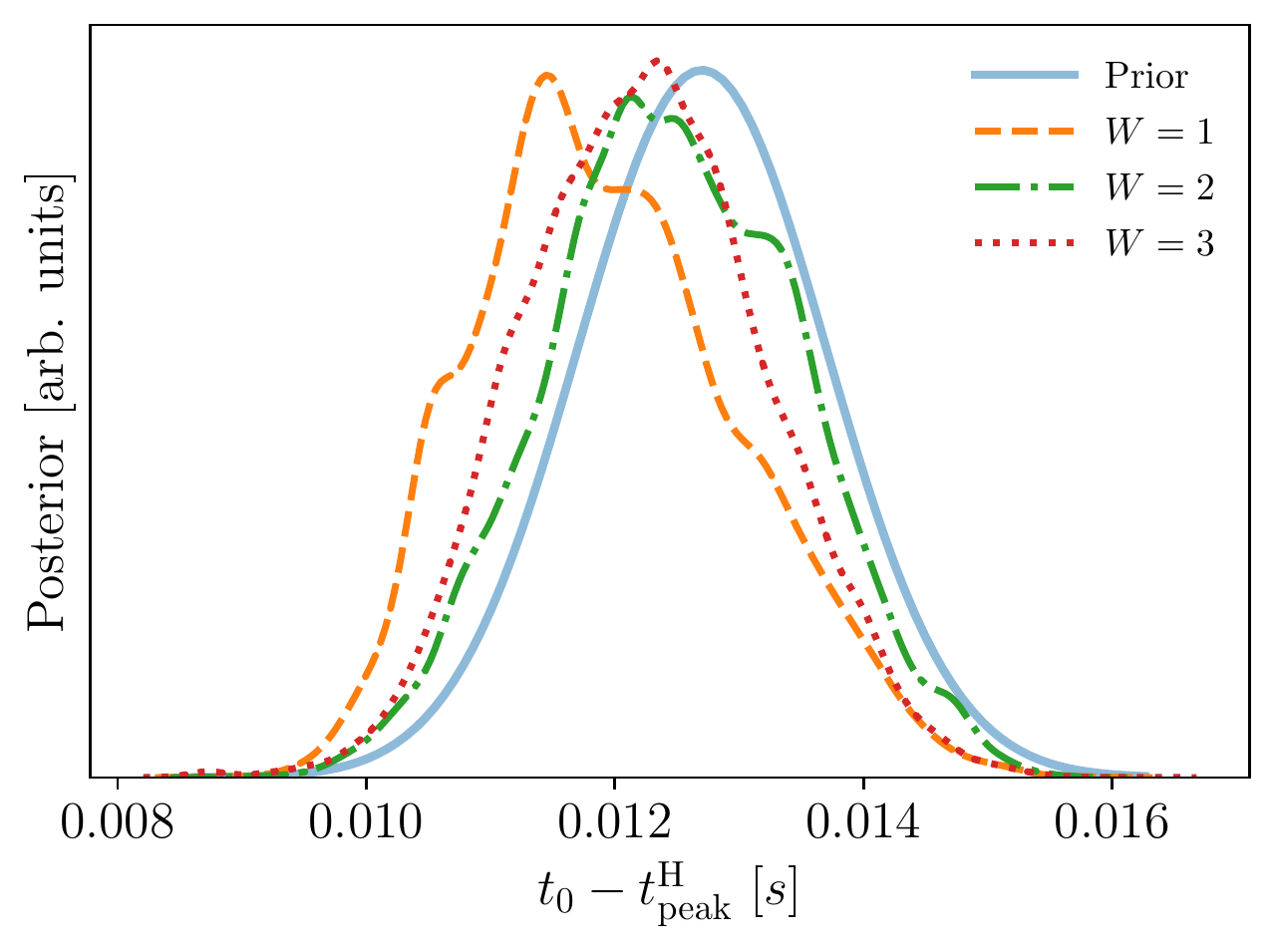}
    \end{minipage}
    \caption{ \label{fig:mass_spin_corner_zero_spin}
    \emph{Left}: 
    Similar to Fig.~\ref{fig:mass_spin_corner_fixed_sky}, posteriors on the remnant mass and dimensionless spin for the GW190521-like injection using a single QNM while marginalizing over the sky position, polarization angle and ringdown start time.
    Also shown in the gray shaded region is the result of a frequency-domain analysis that does not include any wavelets (i.e.\ $W=0$); as expected, since this model has an abrupt discontinuity at $t_0$, this analysis yields severely biased estimates of the remnant mass and spin. This $W=0$ analysis is included here to highlight the important role the wavelets play in our frequency-domain approach.
    \emph{Right}: The prior and posterior distributions on the ringdown start time in the Hanford frame for the same frequency-domain analyses. The prior (solid blue line) is a Gaussian centered $12.7\,\mathrm{ms}$ after the time of peak strain in Hanford, with a standard deviation of $1\,\mathrm{ms}$. The prior has been chosen to be informative; the posterior distributions do not differ significantly from the prior.
    It is necessary to use such an informative prior because we find that the ringdown start time cannot be reliably inferred solely from the data (see discussion in Sec.~\ref{subsec:t0}).
    We observe a slight preference for an early start time when using a small number of wavelets; we speculate that this is due to the wavelet model being less flexible than the maximally flexible model described in Sec.~\ref{subsec:motivation}.
    }
\end{figure*}

First, for reference, an analysis using the time-domain expression for the log-likelihood was performed on this injection searching for the fundamental QNM (i.e.\ $\ell = m = 2$ and $n=0$).
Only the ringdown data $t\geq t_0$ was analyzed; the time series in each interferometer was shifted to geocenter time using the injected sky position, then cut to include $0.1\,\mathrm{s}$ of data from the start of the ringdown.
For each interferometer, the PSD was converted into an autocovariance function using the inverse of the transformation in Eq.~(\ref{eq:WKtheorem}) and this was used to construct the covariance matrix with Eq.~(\ref{eq:Toeplitz}).
We sample over the remnant mass ($M_f$, using a flat prior between $100\,M_\odot$ and $400\,M_\odot$), the dimensionless remnant spin ($\chi_f$, using a flat prior between $0$ and $0.99$), the QNM phase ($\phi_{220}$, using a flat, periodic prior between $0$ and $2\pi$), and the QNM amplitude ($A_{220}$, using a flat prior between 0 and $5 \times 10^{-21}$).
As modes beyond the fundamental will likely have amplitude posteriors consistent with zero, we use a flat prior (as opposed to a log-uniform prior) on the amplitudes; we have checked the choice of prior has minimal influence on the results.
We emphasize that $t_0$ is fixed in this analysis; i.e.\ using a delta-function prior.
The resultant posterior on the remnant parameters is shown in Fig.~\ref{fig:mass_spin_corner_fixed_sky}.
The posterior is consistent with the true remnant properties indicated by the vertical and horizontal lines. 
The true values were obtained with the NRSur3dq8Remnant model \cite{Varma:2018aht, Varma:2019csw, vijay_varma_2018_1435832} which, provided with the injection parameters, can estimate the remnant properties.

Second, the corresponding frequency-domain analyses using $W=1$, 2 and 3 truncated wavelets were also performed on the same injection but now using $4\,\mathrm{s}$ of data centered on the signal.
The same ringdown parameters and priors as before were used. 
In addition, we now sample over the wavelet amplitudes ($\mathcal{A}_w$, using a flat prior between 0 and $5 \times 10^{-21}$), phases ($\varphi_w$, using a flat, periodic prior between $0$ and $2\pi$), widths ($\tau_w$, with flat priors between $5\,\mathrm{ms}$ and $100\,\mathrm{ms}$, or in geometric units between $\sim 4\,M_f$ and $\sim 80\,M_f$) and frequencies ($\nu_w$, with flat priors between $30\,\mathrm{Hz}$ and $100\,\mathrm{Hz}$, or in geometric units between $\sim 0.04\,M_f^{-1}$ and $\sim 0.13\,M_f^{-1}$). 
The label-switching ambiguity among the wavelets was removed by enforcing the ordering $\mathcal{A}_w\leq\mathcal{A}_{w+1}$ via the \emph{hypertriangulation} transformation described in Ref.~\cite{Buscicchio:2019rir}.
We also sample over the wavelet central times ($\eta_w$) using a Gaussian prior with a width of $50\,\mathrm{ms}$ ($\sim 40\,M_f$) centered on the ringdown start time; this choice was empirically found to be sufficiently flexible, whilst also encouraging the wavelets to accurately model the signal near the peak.
Recall that, for the moment, we are fixing the parameters $\alpha$, $\delta$, $\psi$ and $t_0$.
The resultant posteriors on the remnant parameters are shown in Fig.~\ref{fig:mass_spin_corner_fixed_sky}.

From Fig.~\ref{fig:mass_spin_corner_fixed_sky} we see that our frequency-domain approach gives posteriors on the remnant properties that are consistent with both the true values and the time-domain analysis.
We see slight variations in the results of the frequency-domain analyses depending on the number of wavelets used.
This is a high-mass injection with a short inspiral-merger in-band, so it would be expected that a small number of wavelets would be sufficient.
In all cases our frequency-domain approach yields slightly more precise measurements of the remnant properties than the time-domain approach.
We speculate that this is because of some coupling between the ringdown and inspiral-merger parts of the model, which leads to information from the early-time data informing our measurement of the remnant properties.
Indeed, the inspiral-merger model with finite $W$ is only an approximation to the maximally flexible model described in Sec.~\ref{subsec:motivation}.

It is also possible to visually check the performance of the frequency-domain approach by plotting the whitened waveform reconstructions.
These reconstructions, which are the relevant time series that enter the frequency-domain log-likelihood, were found to be in excellent agreement with the injected data. 
Examples of such reconstructions are shown for the GW150914-like injection in Appendix~\ref{app:GW150914}.

\begin{figure*}
    \centering
    \begin{minipage}{0.49\linewidth}
    \includegraphics[width=0.9\textwidth]{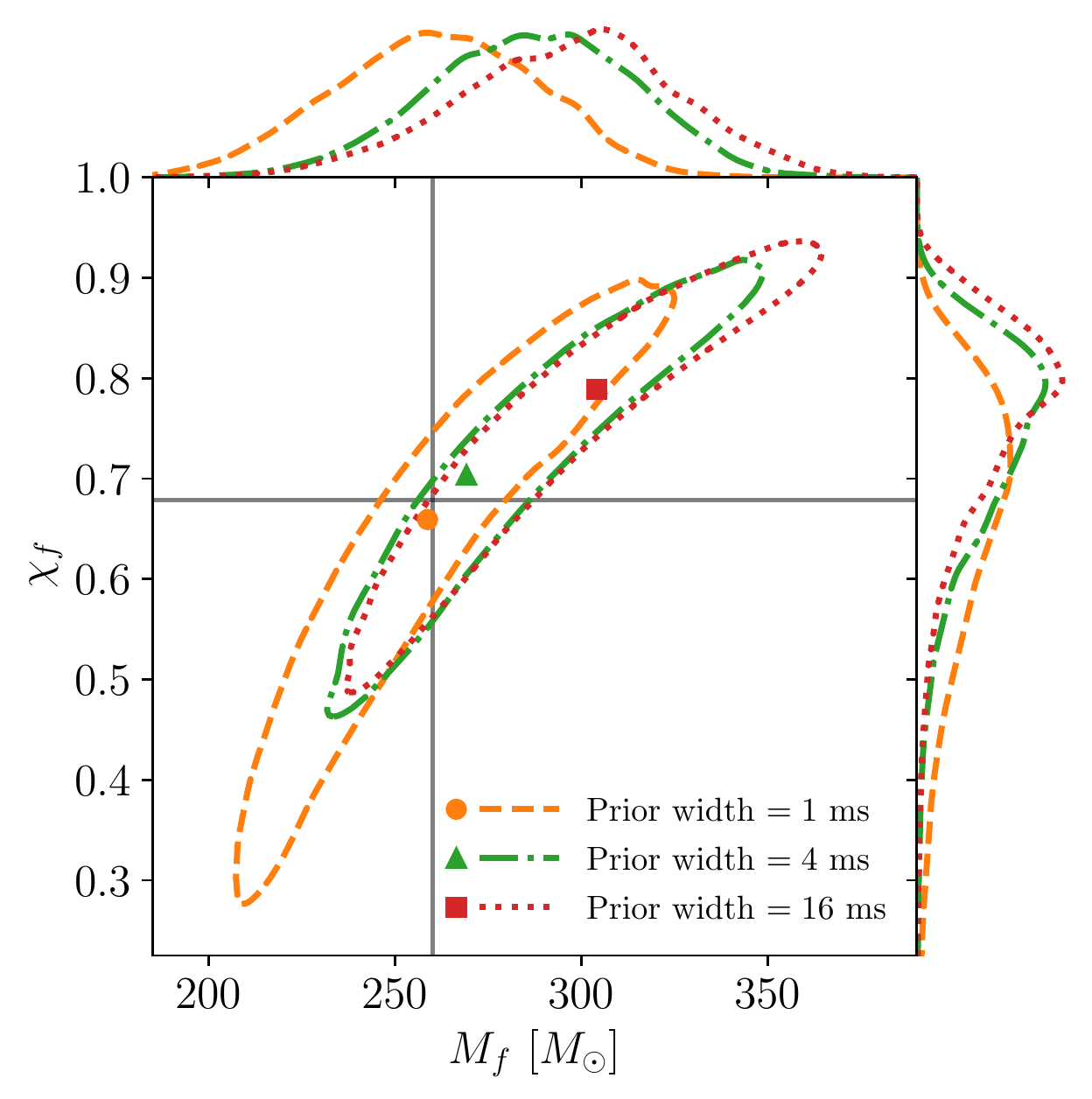}
    \end{minipage}
    \hfill
    \begin{minipage}{0.49\linewidth}
    \vspace{1.2cm}
    \includegraphics[width=0.9\textwidth]{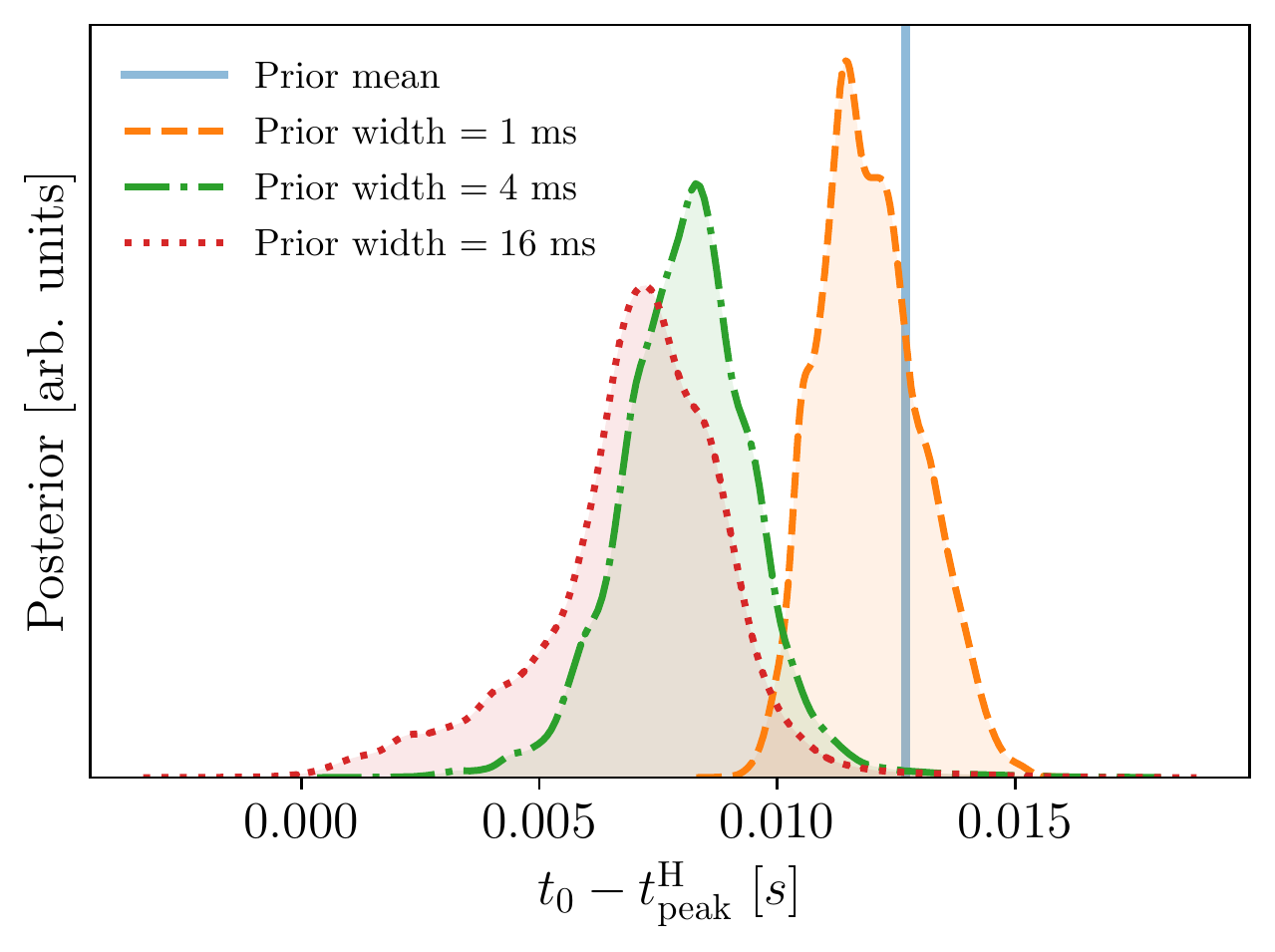} 
    \end{minipage}
    \caption{ \label{fig:start_time_prior}
        \emph{Left}: Similar to Fig.~\ref{fig:mass_spin_corner_zero_spin}, posteriors on the remnant mass and spin for the GW190521-like injection using a single wavelet ($W=1$) and a single QNM using different priors on $t_0$.
        The markers indicate the maximum likelihood values.
        The dashed orange curve is identical to that in Fig.~\ref{fig:mass_spin_corner_zero_spin}.
        \emph{Right}: The corresponding posteriors on the ringdown start time. For each, the prior is a Gaussian centered on the vertical line, with widths given in the legend.
        It can be seen that a wider prior causes earlier ringdown start times to be favored (this is the case even when additional wavelets are included). As a result of the earlier start time, a bias appears in the recovered remnant parameters and higher values of both $M_f$ and $\chi_f$ are favored.
    }
\end{figure*}

We now turn to the case where the source sky position, polarization angle and the ringdown start time are treated as free parameters in the frequency-domain analysis.
We use a uniform prior over the sphere of the sky and a flat, periodic prior on $\psi$ between $0$ and $\pi$.
As the sky position is now allowed to vary in the analysis, the time delay between the different interferometers and the geocenter is not constant. 
Therefore, for the ringdown start time, we choose to place a Gaussian prior on the start time in one of the detectors where the ringdown is clearly visible (we choose Hanford). 
The Gaussian prior was centered on the fixed value used in the time-domain analysis and has a relatively narrow width of $1\,\mathrm{ms}$ ($\sim 0.8\,M_f$).
This choice of prior is quite restrictive (i.e.\ assuming good prior knowledge of $t_0$) and the reasons for this are discussed further in Sec.~\ref{subsec:t0}; although, we note here that this is still more flexible than the delta-function prior used above.
The resultant posteriors on the remnant parameters are shown in left panel of Fig.~\ref{fig:mass_spin_corner_zero_spin}.
The results in the left panel of Fig.~\ref{fig:mass_spin_corner_zero_spin} show that the performance of our frequency-domain approach is not significantly degraded when the searching over $\alpha$, $\delta$, $\psi$ and $t_0$.
Also shown in the right panel of Fig.~\ref{fig:mass_spin_corner_zero_spin} are the posteriors on the ringdown start time, $t_0$ which are discussed in more detail in the next section.

Stochastic sampling was performed using the \texttt{dynesty} \cite{Speagle:2019ivv} implementation of the nested sampling algorithm \cite{doi:10.1063/1.1835238, Skilling:2006gxv}.
For the sampling method, we used random walks with fixed proposals.
Typically, the minimum number of steps used in the random walk was 2000, and the number of live points was 4000.
We note that our frequency-domain ringdown analysis has many more parameters than a time-domain analysis due to the $5W$ parameters used in the wavelet sum.
Posteriors on these inspiral-merger parameters are not presented here, but we note that as the number of wavelets is increased strong degeneracies develop among these parameters.
This is expected, and is desirable in this context, as the wavelet part of the model is designed to be extremely flexible. 
These degeneracies in the inspiral-merger part of the model are not a problem for our present purpose as they do not inhibit our ability to measure the QNMs or the remnant properties. 
All posterior samples, including those for the wavelet parameters, are available at Ref.~\cite{finch_eliot_2021_5569759}.

\subsection{Determining the ringdown start time}\label{subsec:t0}

The ringdown start time, $t_0$, appears as a model parameter in our frequency-domain approach.
This raises two interesting questions: what prior should be placed on $t_0$, and how well can $t_0$ be measured from the data?
The possibility of determining $t_0$ from the data is particularly enticing because the ringdown start time is theoretically uncertain and its choice is crucial for any ringdown analysis.

\begin{figure*}
    \centering
    \includegraphics[width=1.5\columnwidth]{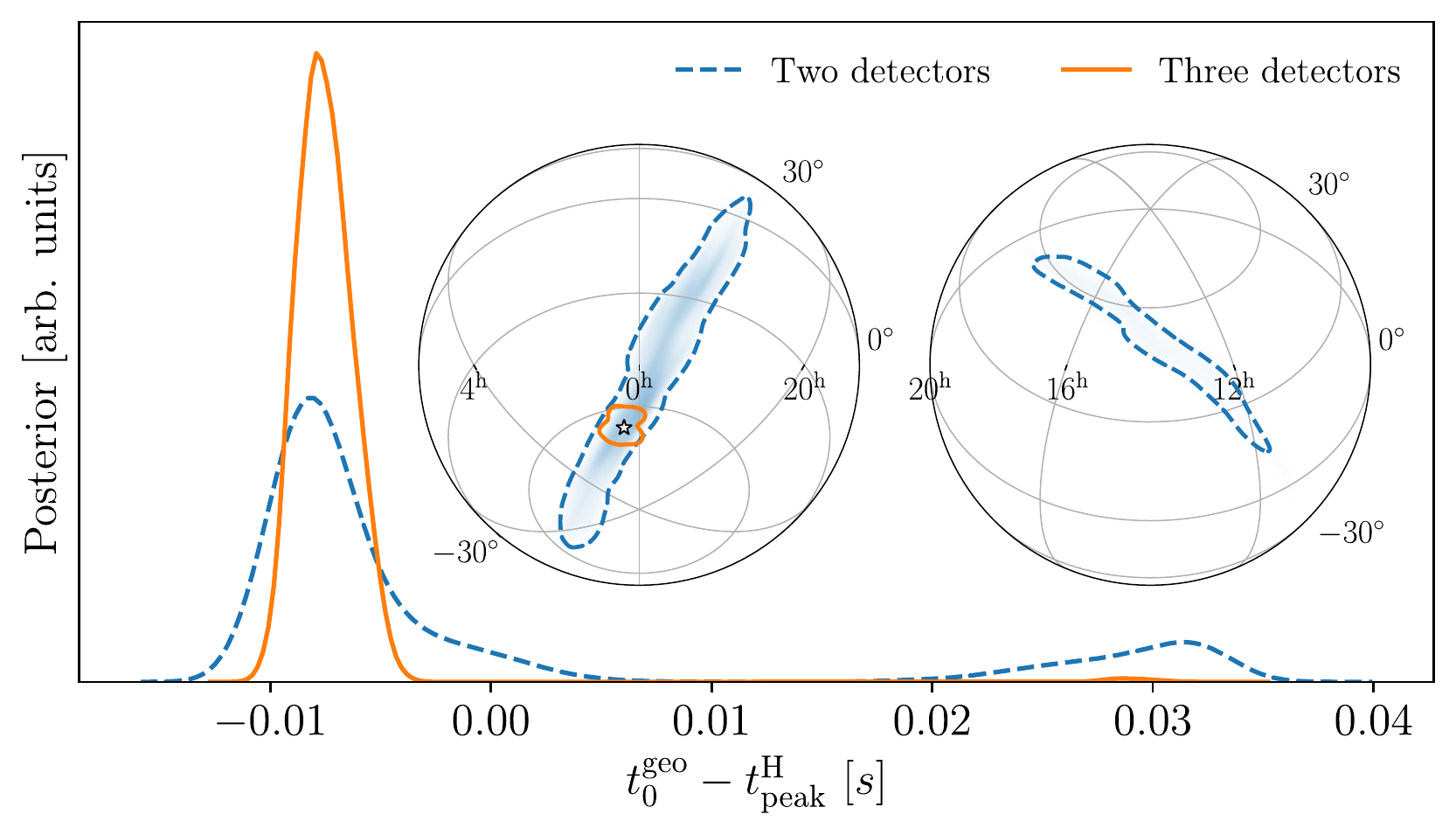}
    \caption{ \label{fig:t0_geocent_posterior}
        \emph{Main panel:} posterior on the ringdown start time in the geocentric frame. The orange line corresponds to the $W=1$ model applied to the three-detector network injection (this is the same analysis presented in Figs.~\ref{fig:mass_spin_corner_zero_spin} and \ref{fig:start_time_prior}, which was plotted with a dashed orange line). 
        The dashed blue line corresponds to a similar analysis on a two-detector network injection (with Virgo removed). This is to motivate the choice to parameterize $t_0$ in the frame of a detector; in the geocentric frame, a multimodal structure appears as a result of different possible sky locations. This makes it harder to place a sensible prior.
        \emph{Left inset plot:} the sky location posterior on the southern hemisphere (orthographic projection). This contains the injected source location (indicated by the star) which is correctly recovered with sky area $\sim 77\ \mathrm{deg}^2$ (90\% confidence) for the three-detector network, and $\sim 1800\ \mathrm{deg}^2$ for the two-detector network.
        \emph{Right inset plot:} the northern hemisphere of the sky contains a secondary mode when using the two-detector network, which correlates with $t_0^\mathrm{geo}$. Both modes of the sky posterior are elongated along the circle of constant time delay between the two detectors.
    }
\end{figure*}

Unfortunately, we find that when using wide priors on $t_0$, early ringdown start times are generally favored and this leads to a bias in the recovered remnant mass and spin.
This can be seen in the results in Fig.~\ref{fig:start_time_prior}, where the $W=1$ analysis previously shown in Fig.~\ref{fig:mass_spin_corner_zero_spin} is repeated with increased values of the $t_0$ prior width.
The posterior on $t_0$ is also affected by the number of wavelets used;
as can be seen from the right panel of Fig.~\ref{fig:mass_spin_corner_zero_spin}, larger values of $W$ tend to favor later ringdown start times. 
We have repeated the analyses in Fig.~\ref{fig:start_time_prior} with larger values of $W$ to see if this counteracts the preference for an early start time (and hence removes biases in the remnant parameters), however, this was found not to be the case.
These calculations show that the posterior obtained on the parameter $t_0$ in our approach depends on the prior and on the number of wavelets used in the (unphysical) model of the inspiral-merger signal.
Therefore, it does not seem to be possible to reliably measure the ringdown start time from the data alone.
It is for this reason that a narrow, informative, $t_0$ prior must be used in the analyses described in the previous section.
Although not desirable, this is still an improvement over the fixed $t_0$ routinely used in most time-domain analyses.

Finally, we discuss the choice that was made in the previous section to place the prior on $t_0$ in the frame of one of the interferometers.
Because the sky position is allowed to vary, using the geocenter time is inappropriate due to coupling with the sky position.
The orange curve in Fig.~\ref{fig:t0_geocent_posterior} shows the posterior on $t_0$ transformed into the geocenter frame from the $W=1$ frequency-domain analysis using the narrow $1\,\mathrm{ms}$ prior on the ringdown start time. 
Also shown in the blue-dashed line is a posterior from an identical injection into a two-interferometer H-L network.
Due to the multimodal sky posterior, the posterior on $t_0$ in the geocenter frame can also be multimodal (this is present in the three-detector analysis to a smaller extent but is most clear in the two-detector analysis).
This makes choosing a suitable prior for the ringdown start time more difficult in the geocenter frame.
It is for this reason that for the analyses described above, the prior was specified in the frame of one of the detectors.
The results in Fig.~\ref{fig:t0_geocent_posterior} also show that our frequency-domain approach yields a posterior on the source sky position as a by-product of the ringdown analysis. However, it should be stressed that this is \emph{not} a ringdown-only result; the entire IMR model, including the unphysical wavelet part, is contributing to this sky localization.

\subsection{Detecting additional QNMs}\label{subsec:overtones}

A key goal in the analysis of BH ringdowns is the detection of additional QNMs beyond the fundamental $\ell=m=2$, $n=0$ mode.
This has already been achieved; see, for example, Ref.~\cite{Isi:2019aib} where the $\ell=m=2$, $n=1$ overtone was identified in GW150914 using a time-domain analysis.
In this section we show, using our GW190521-like injection, that our frequency-domain approach is also able to identify additional QNMs.

\begin{figure*}
    \centering
    \includegraphics[width=0.9\textwidth]{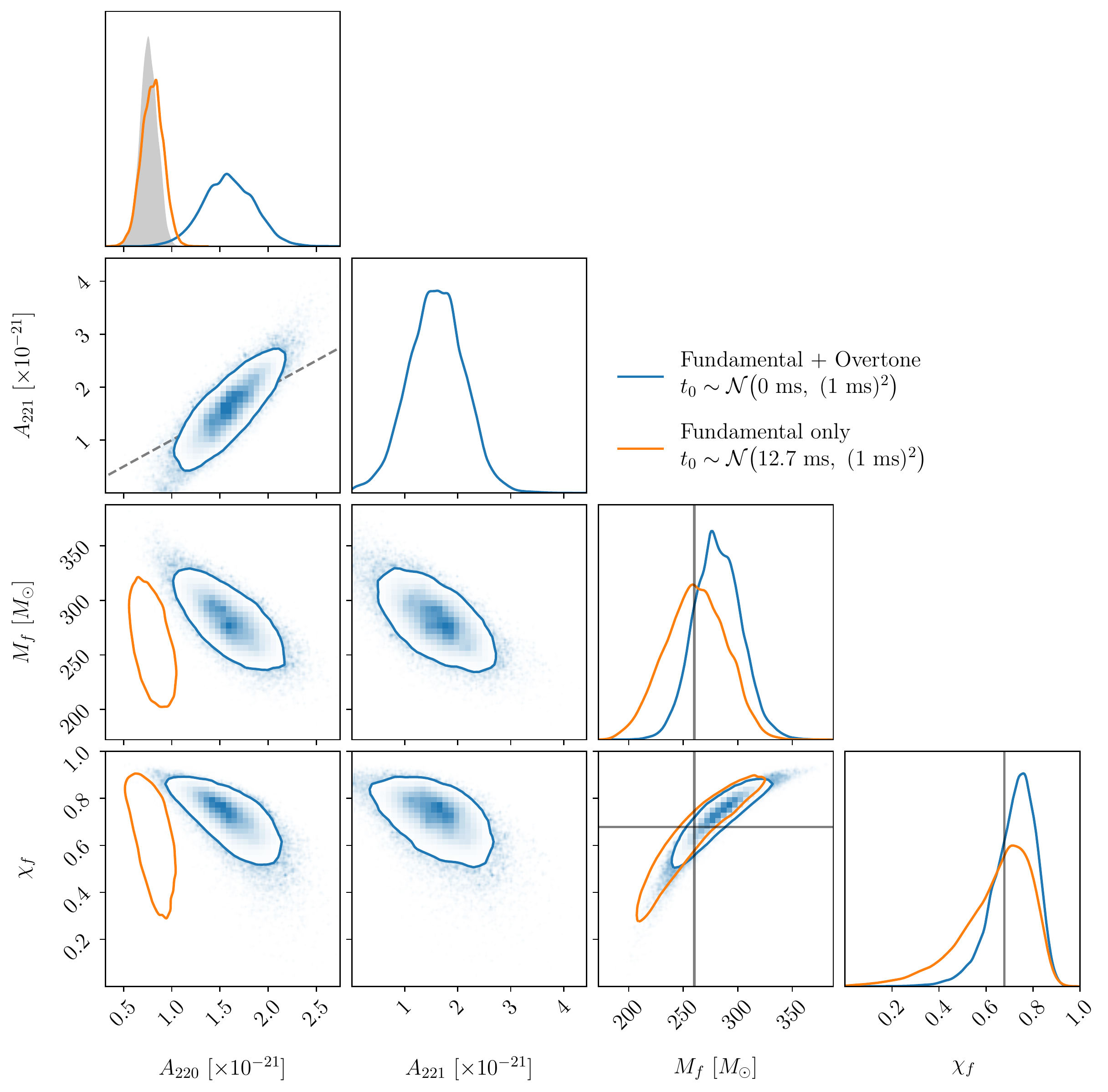}
    \caption{ \label{fig:overtone_corner}
        Posteriors on the QNM amplitudes and remnant mass and spin for one- and two-mode analyses (1QNM and 2QNM respectively) of the GW190521-like injection, performed in the frequency domain.
        The results in blue are for the recovery using two QNMs [the fundamental $(\ell,m,n)=(2,2,0)$ and its first overtone $(2,2,1)$] which are both detected with non-zero amplitudes using a $t_0$ prior centered on the time of the peak strain and with a width of $1\,\mathrm{ms}$.
        Also shown in orange for comparison are the results using one QNM [the fundamental $(2,2,0)$ only] with a prior centered $12.7\,\mathrm{ms}$ after the peak, again with a width of $1\,\mathrm{ms}$.
        The vertical and horizontal solid gray lines indicate the true values of the remnant mass and spin and the diagonal dashed gray line indicates $A_{220}=A_{221}$.
        The difference in the $A_{220}$ amplitude between the two analyses can be explained by the different ringdown start times and the decay of the $\ell = m = 2$, $n = 0$ QNM.
        Over a time $\sim 12.7$ ms, we expect the $A_{220}$ amplitude to decay by a factor $\sim \exp[-12.7\,\mathrm{ms}/\tau_{220}] \approx 0.5$.
        This is shown in the shaded gray posterior in the top-left panel where the results of the 2QNM analysis are used to predict the value of the amplitude at the later start time used by the 1QNM analysis.
    }
\end{figure*}

As a first step towards testing our model we search for the $n=1$ overtone of the fundamental QNM.
It would also be possible to search for higher harmonics (e.g.\ modes with $\ell\geq 3$); however, the results of previous investigations on numerical relativity simulations (see, e.g.\ \cite{Giesler:2019uxc, Ota:2019bzl, Dhani:2020nik, Finch:2021iip}) suggest that overtones are generally more prominent than harmonics in the ringdown and are therefore a natural first target for any search.

We reanalyze the GW190521-like injection in the frequency domain using the $W=1$ inspiral-merger model but this time including an overtone in the ringdown (the choice to use a single wavelet is motivated by the previous results; it is sufficient to model the inspiral-merger for this high-mass injection, and we see no significant improvements with additional wavelets).
When using overtones, it is appropriate to start the ringdown analysis at an earlier time. 
For the frequency-domain analyses a Gaussian prior with a standard deviation of $1\,\mathrm{ms}$ centered at the time of the peak strain was used (this is $12.7\,\mathrm{ms}$ earlier than was used above).
The results of this ``2QNM'' analysis are shown in Fig.~\ref{fig:overtone_corner}, along with the fundamental only ``1QNM'' analysis for comparison.
Posteriors are plotted for the QNM amplitudes and the remnant mass and spin parameters. 
We see that the overtone can be confidently detected with non-zero amplitude.
The 2QNM analysis yields more precise measurements of the remnant mass and spin due to a combination of the earlier ringdown start time (which gives a larger ringdown SNR) and the improved ringdown model.

\begin{figure}[t]
    \centering
    \includegraphics[width=\columnwidth]{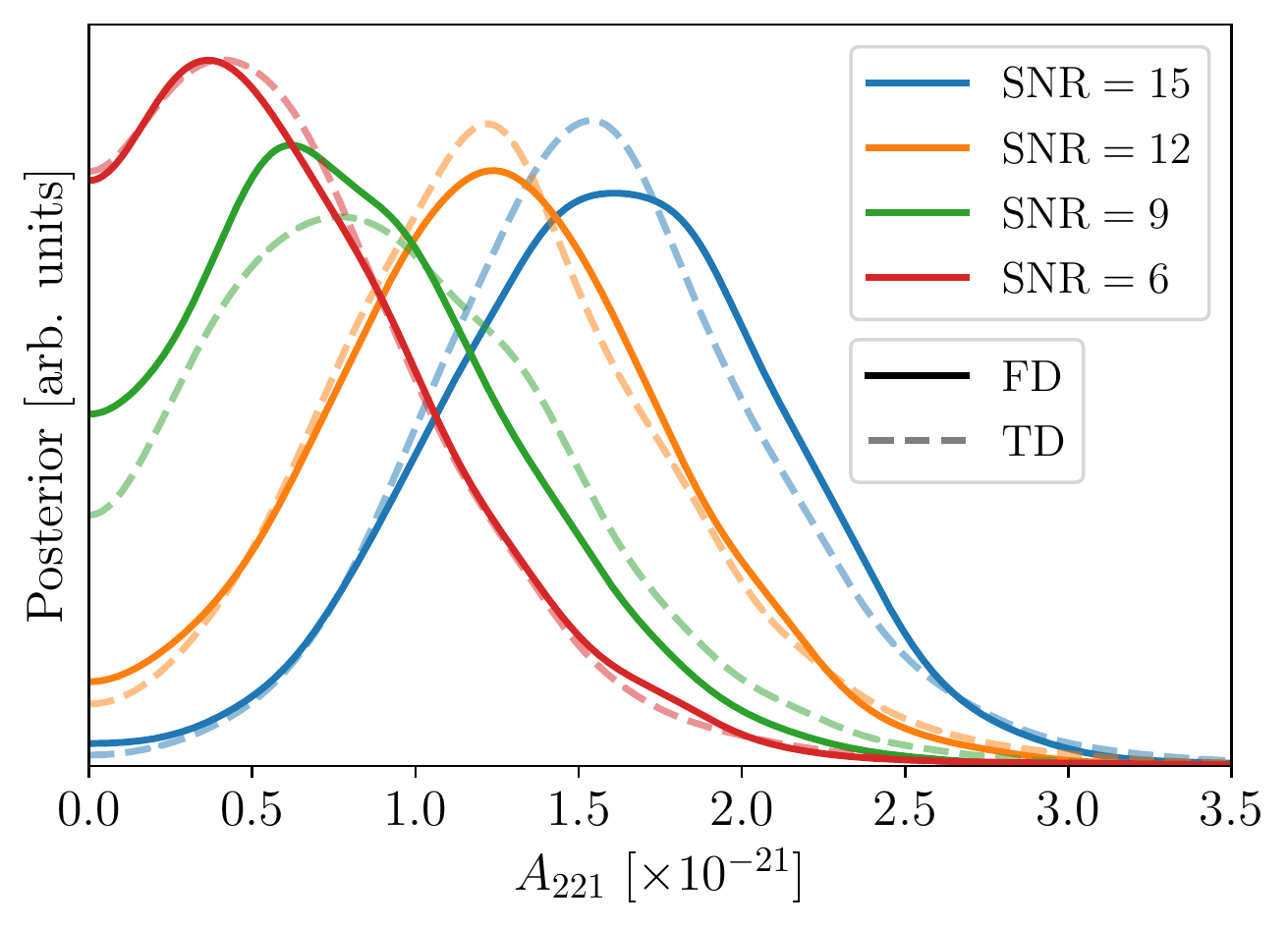}
    \caption{ \label{fig:A221_posterior}
        Overtone amplitude posteriors from a $W=1$ frequency-domain (FD) analysis, where the fundamental $(\ell,m,n) = (2,2,0)$ QNM and its first overtone $(2,2,1)$ are included in the ringdown model. For comparison, the overtone amplitudes from a time-domain (TD) analysis are shown with the dashed lines. The injected SNR in Livingston is controlled by changing the injection luminosity distance: $D_L = \{4016.3,\ 5020.4,\ 6693.9,\ 10040.8\}$ Mpc for $\mathrm{SNR} = \{15,\ 12,\ 9,\ 6\}$ respectively. The maximum likelihood values scale as $D_L^{-1}$.
    }
\end{figure}

We now turn our attention to the resolvability of this additional QNM as a function of the injected SNR and compare the sensitivities of the time- and frequency-domain approaches.
The GW190521-like source was re-injected at a series of lower SNRs: 12, 9, and 6 (in the Livingston detector).
A $W=1$ frequency-domain analysis was re-performed on this sequence of injections, along with a time-domain analysis for comparison. 
Following the use of an earlier ringdown start time in the frequency-domain analysis, for the time-domain analysis the ringdown start time was fixed to the peak of the strain.
The posteriors on the amplitude $A_{221}$ of the overtone are shown in Fig.~\ref{fig:A221_posterior}.
As the SNR is decreased, the overtone becomes increasingly difficult to detect and the posteriors become consistent with $A_{221}=0$.
This is the case for both the time- and frequency-domain analyses which give similar results.
This suggests the time- and frequency-domain approaches are equally sensitive to additional QNMs.

\begin{table}[h]
    \centering
    \begin{tabular}{c|cccc}
    SNR & 15   & 12   & 9   & 6   \\ \hline
    TD & $~ 1.1 ~$ & $0.3$ & $-0.3$ & $-0.7$ \\
    FD & $~ 1.1 ~$ & $~ 0.4 ~$ & $~ -0.4 ~ $ & $~ -0.6 ~$   
    \end{tabular}
    \caption{ \label{tab:bayes_factors}
        The log-Bayes' factors $\log_{10}\mathcal{B}^{2\mathrm{QNM}}_{1\mathrm{QNM}}$ in favor of an overtone for the series of GW190521-like injections at different SNRs, for both time-domain (TD) and $W=1$ frequency-domain (FD) analyses. 
        The uncertainties on these Bayes' factors are all $\pm (0.1$ -- $0.2)$, with errors on the evidences estimated from within a single nested sampling run.
        }
\end{table}

Further evidence supporting this conclusion comes from the odds ratios (aka Bayes' factors) in favor of the overtone.
The Bayes' factors $\mathcal{B}^{2\mathrm{QNM}}_{1\mathrm{QNM}}$ (computed with equal prior odds) in favor of the second QNM were computed from both the time- and frequency-domain analyses. 
In order to do this, we perform an additional set of analyses on the series of injections used in Fig.~\ref{fig:A221_posterior} with the same ringdown start time (fixed at the peak for the time-domain analysis, and a Gaussian prior centered on the peak in Hanford for the frequency-domain) but without the overtone included.
We can then compute the evidence, $\mathcal{B}_{1\mathrm{QNM}}^{2\mathrm{QNM}}$, in favor of the 2QNM analysis (with an overtone) over the 1QNM analysis (fundamental mode only) keeping every other part of the analysis identical. 
The log-Bayes' factors for each of the different SNR injections are shown in Table.~\ref{tab:bayes_factors} where it can be seen that the time- and frequency-domain approaches are equally sensitive to the overtone mode.

\section{Conclusions}\label{sec:discussion}

BH ringdown and QNMs are a key area of study in the burgeoning field of GW astronomy and are particularly important for testing GR.
Ringdown analyses are usually performed in the time domain as this provides a natural way to work with discontinuous models and to apply sharp cuts to the data. 
However, in these analyses the ringdown start time and sky position usually have been fixed beforehand.
The log-likelihood is also more computationally expensive than in the frequency domain.

We have presented a novel approach for analyzing the ringdown in the frequency domain. 
Our approach uses a flexible combination of sine-Gaussian wavelets, truncated at the start of the ringdown, to effectively marginalize over the inspiral and merger parts of the signal. 
The benefits of performing the analysis in the frequency domain include being able to easily vary the source sky position and ringdown start time model parameters as part of the analysis.
As virtually all other GW data analysis is already performed in the frequency domain, a further benefit of our approach is that it allows us to utilize standard, and now very well-tested, GW analysis software packages for performing the Bayesian inference and also for estimating the noise properties.

We have tested our frequency-domain approach by analyzing a series of numerical relativity surrogate injections and by comparing our results with those from a time-domain analysis. 
We find that our frequency-domain approach is equally sensitive to additional QNMs compared to the time-domain approach.
However, we find that it generally yields more precise measurements of the remnant BH mass and spin parameters which we speculate is due to some small coupling with the inspiral and merger signal.
We also paid particular attention to the choice of prior on $t_0$; although this appears as a model parameter in our approach it was found that, unfortunately, it was not possible to reliably determine it solely from the data.

In future we hope to test our method on a larger set of simulated signals, including those with more extreme mass ratios and different spin configurations, and to apply the method to real GW data.


\begin{acknowledgments}
    All posterior samples obtained in this work are made available via Zenodo \cite{finch_eliot_2021_5569759}.
    We thank Riccardo Buscicchio for help producing the sky maps in Fig.~\ref{fig:t0_geocent_posterior}, Davide Gerosa for help implementing the bounded KDEs in Fig.~\ref{fig:A221_posterior}, and Will Farr, Maximiliano Isi, Gregorio Carullo and other members of the LVK testing GR group for useful discussions.
    Computations were performed using the University of Birmingham's BlueBEAR HPC service.
\end{acknowledgments}


\bibliographystyle{apsrev4-1}
\bibliography{bibliography}


\appendix
\section{GW150914-like Injection}\label{app:GW150914}

In the main body of the paper the frequency-domain ringdown analysis was tested on GW190521-like injections with varying SNR and observed using a network of two or three interferometers. It was found to perform well. In this appendix we test the frequency-domain approach further by analyzing a GW150914-like injection.

The surrogate was initialized with a total mass of $72.2\,M_\odot$ and a mass ratio of $1.16$. 
As before, all of the component spins were set to zero for simplicity. 
The simulated sky location and GW polarization angle were taken to be $\alpha = 1.95$, $\delta = -1.27$, and $\psi = 0.82$. These are consistent with the GW150914 posterior and were chosen to match the values used in \cite{Isi:2019aib}.
The distance to the binary was set to $471.4\,\mathrm{Mpc}$, which gave an optimal SNR in Hanford of 25.

The inclination angle was chosen to be $\pi$ (i.e\ the source is injected ``face-off'') which is consistent with the GW150914 posterior.
The source inclination affects the GW polarization, and this is handled via the introduction of a ``ellipticity parameter'' $\epsilon$ which has the effect of transforming $h_\times(t) \rightarrow \epsilon h_\times(t)$.
For the ``face-on'' injections in the main text $\epsilon=1$ was used, while for the ``face-off'' injections considered here $\epsilon=-1$.
A more general analysis would allow $\epsilon$ to vary as a free parameter, such as what was done in Ref.~\cite{Isi:2021iql}.

We perform zero-noise injections into the two-interferometer H-L LIGO network that was operating at the time of the first detection. 
We use the PSDs associated with the data surrounding GW150914 (available at Ref.~\cite{gwtc1psds}).
These different parameters (particularly the lower total mass) result in a signal with a longer inspiral.
This is an important test case for our model as there is a much larger fraction of the SNR in the inspiral-merger (compared to the GW190521-like injection) which has to be ``marginalized out'' in the analysis.

As was done initially for the GW190521-like injection, we fix the sky location and polarization angle to the injected values to simplify the problem and aid comparison to the time-domain analysis.
The ringdown start time is also fixed to $3\,\mathrm{ms}$ after the time of the peak strain ($\sim 10\,M_f$ in geometric units).

\begin{figure}[b]
    \centering
    \includegraphics[width=\columnwidth]{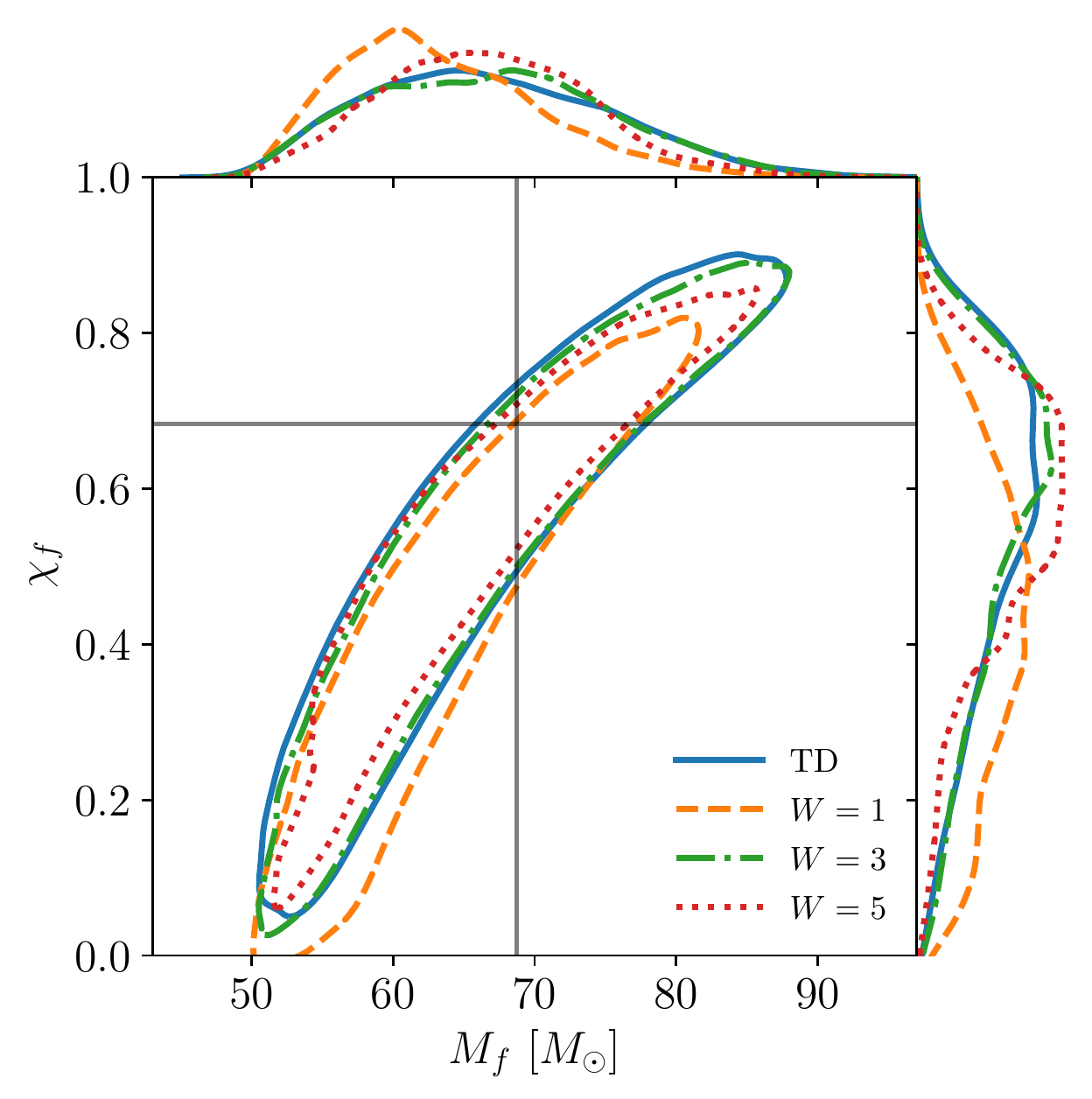}
    \caption{ \label{fig:GW150914_mass_spin_corner}
    Similar to Fig.~\ref{fig:mass_spin_corner_fixed_sky}, posteriors on the recovered remnant mass and spin for the GW150914-like injection using the fundamental QNM and a varying numbers of wavelets. 
    Also shown for comparison is the result of a time-domain analysis (solid blue line).
    }
\end{figure}

Following the same procedure as in Sec.~\ref{sec:injection_study}, a time-domain analysis was first carried out to recover the fundamental QNM. 
The prior on the remnant mass was adjusted to reflect the lower injected value (flat between $50\,M_\odot$ and $100\,M_\odot$), but otherwise the analysis was unchanged from the time-domain analyses described in the main text.
The resultant remnant mass and spin posterior is shown by the blue solid line in Fig.~\ref{fig:GW150914_mass_spin_corner}.

\begin{figure*}[t]
    \centering
    \includegraphics[width=2\columnwidth]{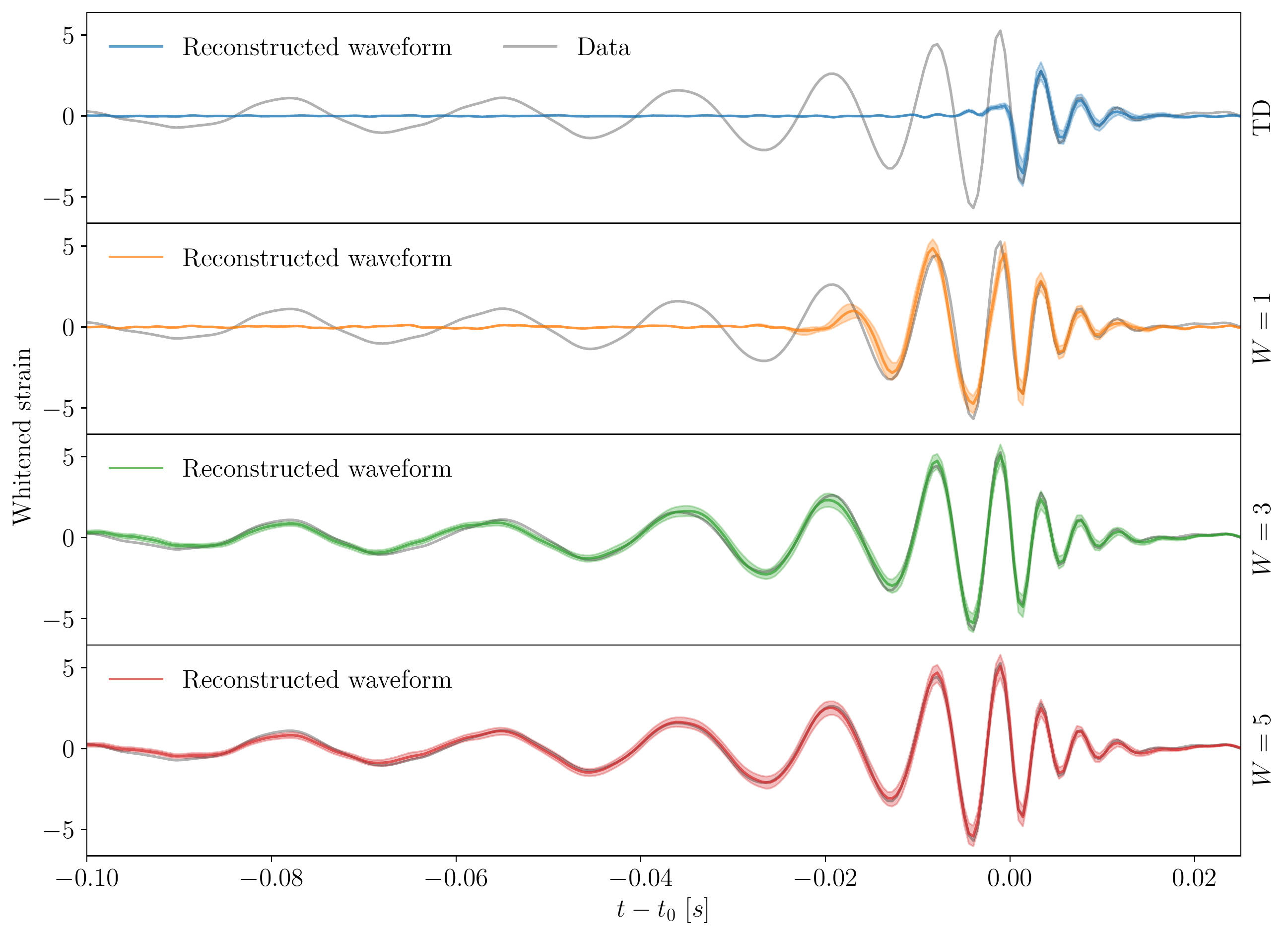}
    \caption{ \label{fig:GW150914_whitened_waveforms}
    Whitened waveform reconstructions (in Hanford) corresponding to the results of Fig.~\ref{fig:GW150914_mass_spin_corner}. 
    The top panel shows the waveform from a time-domain analysis. 
    In the time-domain approach the data before the ringdown start time is excluded, but here the waveform is plotted for all times.
    This highlights one of the problems with using a ringdown-only model in the frequency domain: the abrupt start of the model leads to spectral leakage when Fourier transforming (visible as oscillations before the ringdown start time).
    The following panels show waveforms from the frequency-domain approach.
    Problems with spectral leakage are avoided, due to the wavelets smoothly connecting to the ringdown part of the model.
    Just a single wavelet fails to model the full GW150914-like inspiral-merger, which is to be expected because of its longer duration in-band.
    As more wavelets are included in the model, more of the inspiral-merger is captured by the model.
    The difference in the reconstruction for three and five wavelets is minimal, showing the model is converging on the signal.
    }
\end{figure*}

Secondly, a series of frequency-domain analyses were carried out using an increasing number of wavelets.
Results for $W=1$, 3, and 5 are shown in Fig.~\ref{fig:GW150914_mass_spin_corner}.
We found a slightly more restrictive prior on the wavelet central times, $\eta_w$, was required to aid the inference; the Gaussian width was reduced to $10\,\mathrm{ms}$ ($\sim 30\,M_f$).
This encouraged the wavelets to fit near the ringdown start time, which is the part of the signal we are most interested in.
The upper bound on the wavelet frequencies was also increased to $500\,\mathrm{Hz}$ ($\sim 0.17\,M_f^{-1}$), as the lower binary mass means the merger-ringdown occurs at a higher frequency.
We see that a single wavelet is not quite sufficient to avoid bias in the remnant parameters, which may be expected when working with a longer inspiral.
As the number of wavelets is increased the bias disappears, and the remnant posteriors seem to converge to a solution that is stable against the inclusion of additional wavelets. 
As was the case for the GW190521-like analysis, we find the frequency-domain model achieves tighter constraints on the remnant parameters in comparison to the time-domain analysis. 

Finally, we inspect the whitened waveform reconstructions for all four analyses shown in Fig.~\ref{fig:GW150914_mass_spin_corner}.
We focus on the waveform in the Hanford detector.
We take samples from the posterior of each run and use these to compute the projected waveform $F^\mathrm{H}_{+} h_+(t+\Delta t_\mathrm{H})+F^\mathrm{H}_{\times} h_\times(t+\Delta t_\mathrm{H})$, see Eq.~(\ref{eq:projection_antenna}). This quantity is always discontinuous for both the time- and frequency-domain analyses.
We then whiten this waveform (and the data) using the Hanford PSD. After whitening, the projected waveform is continuous.
In Fig.~\ref{fig:GW150914_whitened_waveforms} we plot the median and $5\%-95\%$ credible region of the whitened waveform reconstructions.
The figure highlights the problem of spectral leakage, which occurs when taking Fourier transforms of discontinuous models. 
The time-domain model (top panel) has a discontinuity at the ringdown start time and this causes oscillations to appear before the start time in the whitened waveform.
The following panels, which include wavelets to model the inspiral-merger signal, remove this discontinuity and prevent Fourier transform artifacts.
A single wavelet is not sufficient to capture the full inspiral-merger signal, which likely causes the bias seen in Fig.~\ref{fig:GW150914_mass_spin_corner}. 
Increasing the number of wavelets makes the model flexible enough to model the inspiral-merger signal, and also to remove bias in the mass-spin posterior.

\end{document}